\documentclass{llncs}

\usepackage{graphicx}
\usepackage{amsmath,amssymb,amsfonts} 
\usepackage[varg]{txfonts}
\usepackage{caption}
\usepackage{color}
\usepackage{array,booktabs}
\usepackage{tabularx, booktabs}
\usepackage{supertabular,booktabs}
\usepackage{rotating}
\usepackage{multirow}
\usepackage[T1]{fontenc}
\usepackage{url}
\usepackage{pifont}
\usepackage{adjustbox} 
\usepackage{enumerate}
\usepackage{mathrsfs}
\usepackage{float}
\usepackage{enumerate}
\usepackage{enumitem}
\usepackage{longtable}
\usepackage{tikz}
\usetikzlibrary{positioning}
\usetikzlibrary{shapes,arrows,automata,positioning}
\usepackage{pgfplots}
\usepackage{thmtools,thm-restate}

\usepackage[linesnumbered,ruled,vlined]{algorithm2e}
\newcolumntype{M}[1]{>{\centering\arraybackslash}m{#1}}
\newcolumntype{Y}{>{\centering\arraybackslash}X}
\SetKwRepeat{Do}{do}{while}

\usepackage{url}
\urldef{\mailsa}\path|jakub.breier@gmail.com, ho0001lu@e.ntu.edu.sg, djap@ntu.edu.sg, malei@hit.edu.cn,sbhasin@ntu.edu.sg, yangliu@ntu.edu.sg|      
\newcommand{\keywords}[1]{\par\addvspace\baselineskip
\noindent\keywordname\enspace\ignorespaces#1} 

\DeclareGraphicsExtensions{.pdf}

\begin{document}

\title{DeepLaser: Practical Fault Attack on Deep Neural Networks}
\author{Jakub Breier$^1$ \and Xiaolu Hou$^2$ \and Dirmanto Jap$^1$ \and Lei Ma$^{2,3}$ \and Shivam Bhasin$^1$ \and Yang Liu$^2$}
\authorrunning{J. Breier, et. al}

\institute{$^1$Physical Analysis and Cryptographic Engineering\\Temasek Laboratories at Nanyang Technological University, Singapore\\
$^2$School of Computer Science and Engineering\\
Nanyang Technological University, Singapore\\
$^3$School of Computer Science and Technology\\
Harbin Institute of Technology, China\\
\mailsa\\
}

\maketitle

\begin{abstract}
As deep learning systems are widely adopted in safety- and security-critical applications, such as autonomous vehicles, banking systems, etc., malicious faults and attacks become a tremendous concern, which potentially could lead to catastrophic consequences.
In this paper, we initiate the first study of leveraging physical fault injection attacks on Deep Neural Networks (DNNs), by using laser injection technique on embedded systems.
In particular, our exploratory study targets four widely used activation functions in DNNs development, that are the general main building block of DNNs that creates non-linear behaviors -- ReLu, softmax, sigmoid, and tanh.
Our results show that by targeting these functions, it is possible to achieve a misclassification by injecting faults into the hidden layer of the network.
Such result can have practical implications for real-world applications, where faults can be introduced by simpler means (such as altering the supply voltage).
\keywords{deep learning security, fault attacks, adversarial attacks}
\end{abstract}

\section{Introduction}

Internet of things (IoT) and artificial intelligence (AI) are the two integral components of modern paradigms like smart city, self-driving cars etc.
Recent advancement in AI has made it applicable to complex scenarios.
The most efficient AI is known in the form of deep learning which has achieved great leaps of progress in many application domains, such as computer vision, robotics, gaming, finance, medical systems~\cite{DNNprogress}.
In parallel, IoT has pushed the computing elements (and sensors) outside traditional boundaries and motivates to place them everywhere.
This has also enabled easy physical access to the computing elements which was not possible previously.

We look at a specific class of physical attacks known as fault attacks, which has become a reality owing to decreasing price and expertise required to mount such attack.
Fault attacks are active attacks on a given implementation which try to perturb the internal software computations by external means.
The adversary uses methods like voltage glitches or laser injection to introduce perturbations for various purposes, ranging from erroneous computation, denial of service etc.
Such attacks are commonly used for mounting secret key recovery attacks in cryptography or violate/bypass security checks~\cite{joye2012fault}.
In this work, we analyze deep learning under fault attacks.

Deep learning is the family of neural networks composed of an input layer, three or more hidden layers and an output layer.
Based on the internal structure, several candidates exist like multi-layer perceptron (MLP), convolutional neural networks (CNN), recurrent neural network (RNN) etc.
These are popularly known as deep neural networks (DNN).
While each of these architectures has unique functions, we focus on activation functions which remain common across architectures and are an important part of the algorithm to obtain non-linear behaviors~\cite{Goodfellow-et-al-2016}.
These commonly used activation functions are: \texttt{softmax}, \texttt{ReLu}, \texttt{sigmoid} and \texttt{tanh}.
Studying these functions under fault attacks allows to derive general conclusions on susceptibility of deep learning to fault attacks.

We implemented the most common activation functions used across DNNs on a low-cost microcontroller (often used in IoT).
Next, we performed practical laser fault injection using a near-infrared diode pulse laser to inject faults during the processing of activation function. The use of laser facilitates a strong attacker model with extensive fault injection capabilities.
With the models, derived from practical fault injection, we analyze the susceptibility of DNN against such attacks.
The primary goal of the performed attacks is to achieve misclassification during the testing phase.
In the hindsight, the achieved misclassification can jeopardize the functioning of DNN-based paradigms like smart city.

Extensive studies have been performed on adversarial attacks, that crafts the input data with little perturbation to fool deep learning systems~\cite{goodfellow2014explaining,Heboundaryanalysis,szegedy2014intriguing,carlini2017towards,carlini2017adversarial,iclr18aboundaryattack,iclr18aespatial}.
To the best of our knowledge, our study is the first work to explore practical~(physical) fault injection on deep neural network, where we focus on attacking the DNNs itself instead of creating input data to fool DNNs like adversarial attack does.

\section{Background}
\subsection{Fault Injection Attacks}
Fault injection attacks are a popular physical attack vector used against cryptographic circuits~\cite{Barthe14}.
By changing intermediate values during the cryptographic algorithm execution, they can efficiently provide information on secret values, helping to recover the secret key in just a few encryptions~\cite{biham1997differential,boneh1997importance,breier2015laser}. 
Normally, the secret key recovery would require infeasible amount of computing time.
Similarly, these attacks can be used against verification circuits, such as \texttt{PIN} verification on a smartcard, where a comparison function can be skipped and grant access to a malicious user~\cite{giraud2004survey}.

When it comes to fault injection techniques, there are several options one can use, mostly depending on the adversary budget and expertise~\cite{bar2006sorcerer}.
The most basic methods include variations in voltage or clock signal, allowing disturbance of instruction sequences in microcontrollers~\cite{balasch2011depth}.
Electromagnetic fault injection allows more precise location targeting, enabling faults in memories~\cite{kim2007faults,moro2013electromagnetic}.
Laser fault injection is the most precise from commonly used techniques, being capable of flipping single bits~\cite{agoyan2010flip}.

Up to date, to the best of our knowledge, only \cite{liu2017fault}
describes fault injection attack on neural networks.
In their paper, they only provide a white box attack on deep neural network through software simulation, while observing the changes in the output after introducing faults in the network's values.
However, they do not provide insight on practicality of such attack. 
Whether such attacks could also be applied physically remained an open problem.
Therefore, in our paper, we experimentally show what types of faults are achievable in practice and we further use this information to develop a realistic attack on DNNs.

\subsection{Activation Functions}
The activation functions we consider are the following: \texttt{softmax}, \texttt{ReLu}, \texttt{sigmoid} and \texttt{Tanh} \cite{Goodfellow-et-al-2016}.

Softmax is normally used as the activation function for output layer.
It takes a vector $\mathbf{x}$ as input, $i$th entry of the output gives the probability of a given input belonging to class $i$:
\begin{equation*}
\text{softmax}(x)_i=\frac{\text{exp}(x_i)}{\sum_{j}\text{exp}(x_j)},
\end{equation*}
where exp is the exponentiation function with base $e$.

In modern neural networks, the default recommendation for activation function is the rectified linear unit or ReLu defined as follows:
\begin{equation*}
    \text{ReLu}(x)=\max\{0,x\}.
\end{equation*}
It is a piecewise linear function which preserves properties that make the optimization of linear model easy.

Before the introduction of ReLu, commonly used activation functions are logistic sigmoid activation function
\begin{equation*}
    \text{sigmoid}(x)=\frac{1}{1+\text{exp}(-x)},
\end{equation*}
and hyperbolic tangent function
\begin{equation*}
    \text{tanh}(x)=\frac{2}{1+\text{exp}(-2x)}-1.
\end{equation*}

The sigmoid function is normally used to introduce non-linearity in the model.
A reason for its popularity comes from the simple equation between its derivative and itself: $\text{sigmoid}'(x)=\text{sigmoid}(x)\allowbreak(1-\text{sigmoid}(x))$.
However, sigmoid functions becomes insensitive to inputs with large absolute values.
In such cases, the hyperbolic tangent activation function is used as an alternative.


\section{Practical DNN Attack Feasibility Analysis}
\label{sec:experiment}
In this part we first show the practical laser fault attack setup in Section~\ref{sec:equipment}.
In Section~\ref{sec:attack} we show the possible fault attacks on activation functions that we have discovered with practical experiments.
In Section 4, those attacks will be used for simulating missclassification attacks on MNIST DNNs.
\subsection{Attack Equipment Setup}
\label{sec:equipment}
The main component of the experimental laser fault injection station is the diode pulse laser.
It has a wavelength of 1064 nm and pulse power of 20 W. 
This power is further reduced to 8 W by a 20$\times$ objective lens which reduces the spot size to 15$\times$3.5 $\mu m^2$.

\begin{figure}[tb]
    \centering
    \begin{tabular}{cc}
    \includegraphics[width=0.45\textwidth,angle=-90,origin=c]{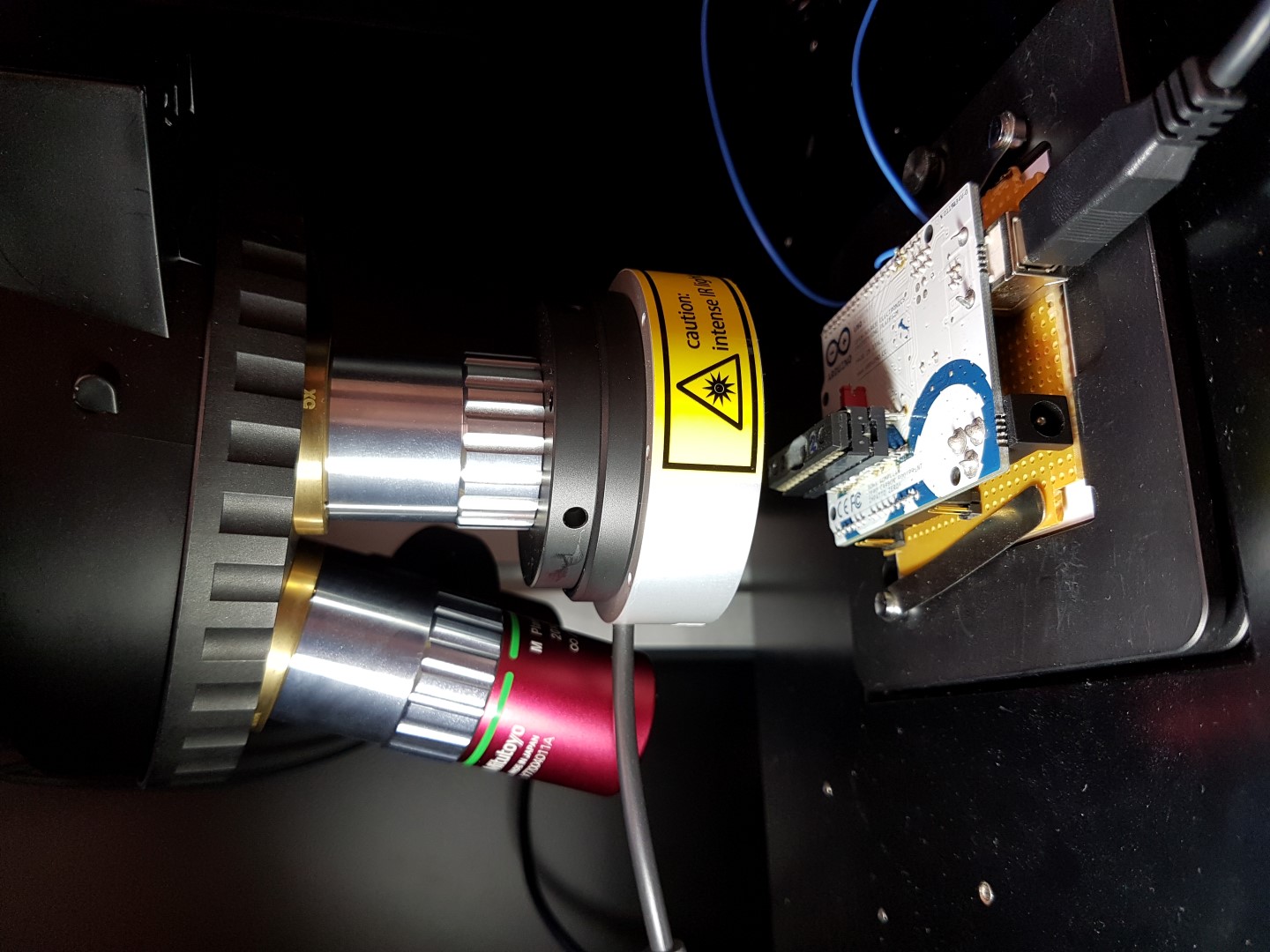}&
    \includegraphics[width=0.54\textwidth]{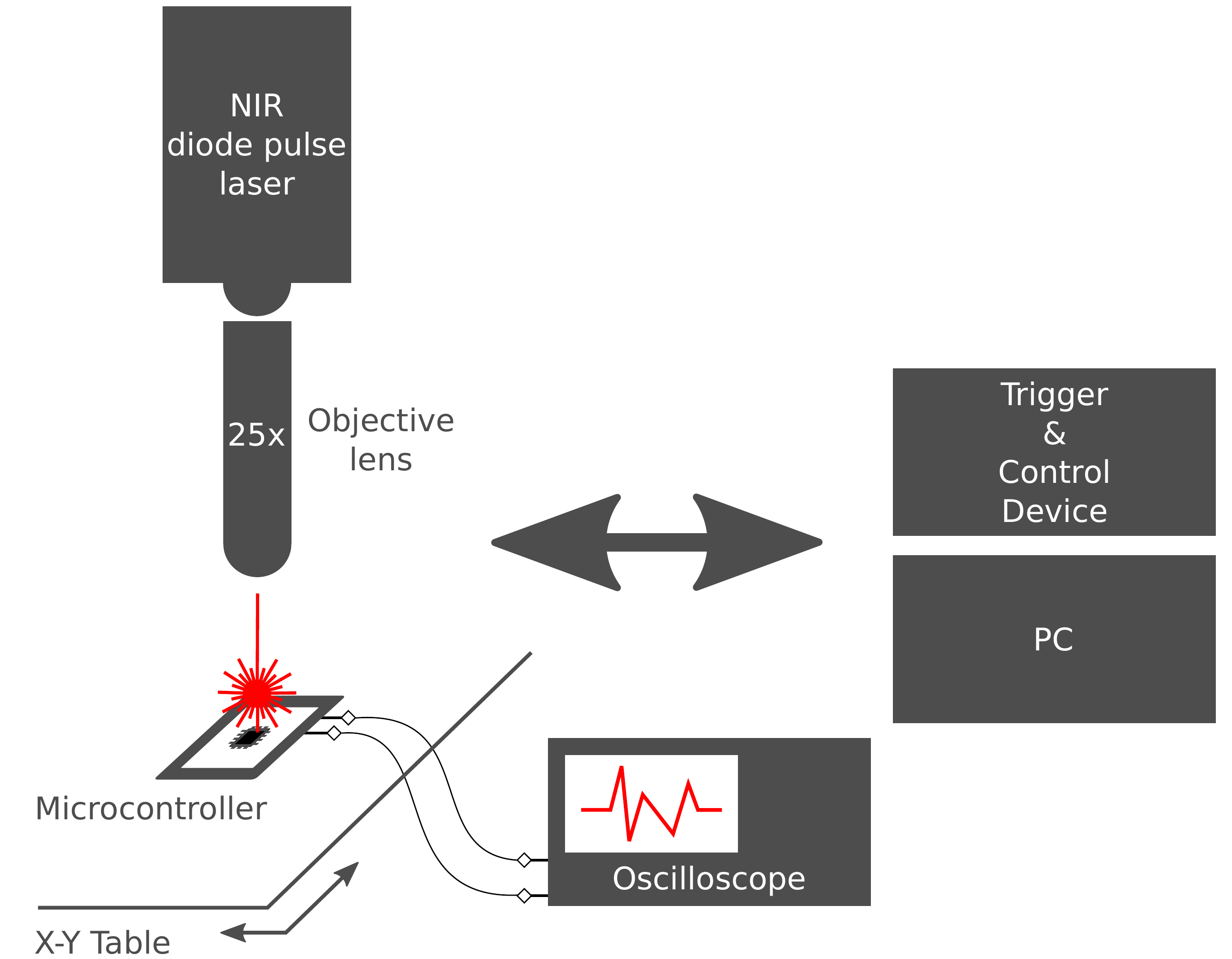}\\
    (a) & (b)
    \end{tabular}
    
    \caption{Experimental laser fault injection setup -- (a) device under test, (b) setup components.}
    \label{fig:setup}
\end{figure}

As the device under test (DUT), we used \texttt{ATmega328P} microcontroller, mounted on \texttt{Arduino UNO} development board.
The package of this chip was opened so that there is a direct visibility on a back-side silicon die with a laser.
The board was placed on an \texttt{XYZ} positioning table with the step precision of 0.05 $\mu m$ in each direction.
A trigger signal was sent from the device at the beginning of the computation so that the injection time could be precisely determined.
After the trigger signal was captured by the trigger and control device, a specified delay was inserted before laser activation.
Laser activation timing was also checked by a digital oscilloscope for a greater precision.
Our setup is depicted in Figure~\ref{fig:setup}.

The chip area is 3$\times$3 mm$^2$ large, while the area sensitive to laser is $\approx$ 50$\times$70 $\mu m^2$ large.
With a laser power of 4.5\% we were able to disturb the algorithm execution, when tested with reference codes.

\subsection{DNN Activation Function Fault Analysis}
\label{sec:attack}
To evaluate different activation functions, we implemented three simple $3$-layer neural networks with sigmoid, ReLu and tanh as the activation function for the second layer respectively. The activation function for the last layer was set to be softmax.
The neural networks were implemented in C programming language, which were further compiled to AVR assembly and uploaded to the DUT.

We surrounded the activation functions in the second layer with a trigger signal that raised a voltage on a selected Arduino board pin to 5 V, helping us to determine the proper laser timing.

As instruction skip/change are one of the most basic attacks on microcontrollers, with high repeatability rates~\cite{breier2015laser}, we aimed at this fault model in our experiments.
The microcontroller clock is 16 MHz, one instruction takes 62.5 ns. 
Some of the activation functions took over 2000 instructions to execute.
To check what are the vulnerabilities of the implementations, we have carefully varied the timing of the laser glitch from the beginning until the end of the function execution so that every instruction would be eventually targeted.

Please note that we used a single fault adversarial model, meaning that exactly one fault was injected during one activation function execution.

\begin{figure}[tb]
    \centering
    \includegraphics[width=0.6\textwidth]{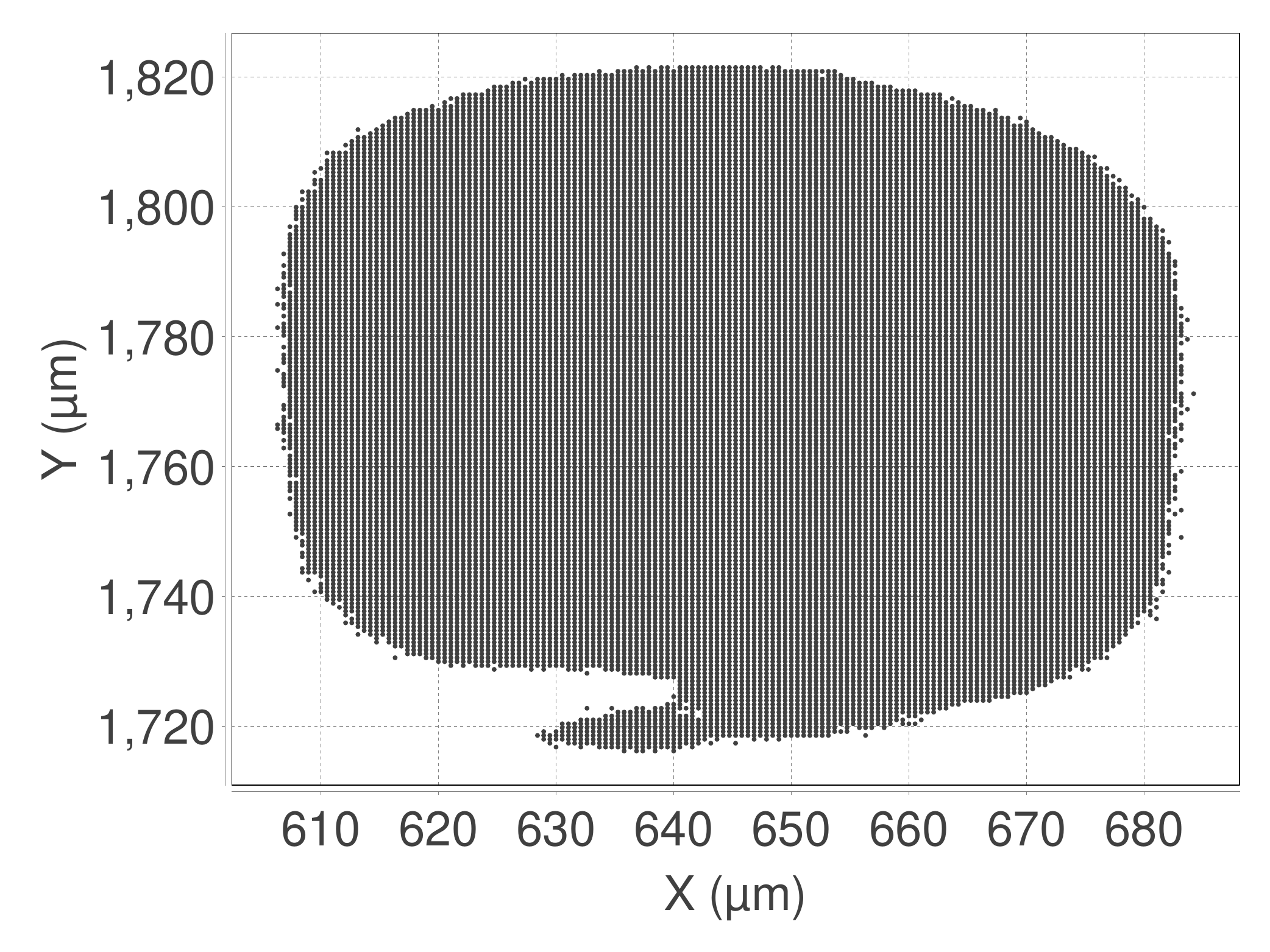}
    \caption{Area plot depicting successful instruction skip experiments.}
    \label{fig:area}
\end{figure}

After we observed a successful misclassification, we determined the vulnerable instructions by visual inspection of the compiled assembly code and by checking the timing of the laser in that particular fault injection instance.
Area of the chip vulnerable to these disturbances is depicted in Figure~\ref{fig:area}.

In this exploratory study, we implemented a random neural network, consisting of 3 layers, with 19, 12, and 10 neurons in input layer, hidden layer, and output layer, respectively.
Our fault attack was always targeting the computation of one of the activation functions in hidden layer.
In the following, we will explain the experimental results on different activation functions in detail.

\noindent
\textbf{ReLu.} This function is implemented by a following code in C:

\noindent
   \hspace*{50pt}\texttt{if (Accum > 0) \{}\\
   \hspace*{60pt}\texttt{       HiddenLayerOutput[i] = Accum;\}} \\
   \hspace*{50pt}\texttt{else \{}\\
   \hspace*{60pt}\texttt{       HiddenLayerOutput[i] = 0;\}}\\ 
where $i$ loops from $1$ to $12$ so that each loop gives one output of the hidden layer.
\texttt{Accum} is an intermediate variable that stores the input of activation function for each neuron.

The assembly code inspection showed that the result of successful attack was executing the statement after else such that the output would always be 0.
The corresponding assembly code is as follows:
\begin{center}
\setlength{\tabcolsep}{3pt}
\begin{tabular}{rlll}
\textcolor{gray}{1}&   & \texttt{ldi r1, 0} & ;load 0 to r1\\
\textcolor{gray}{2}&   & \texttt{cp r1, r15} & ;compare MSB of \texttt{Accum} to \texttt{r1}\\
\textcolor{gray}{3}&   & \texttt{brge else} & ;jump to \texttt{else} if $0 >=$ \texttt{Accum}\\
\textcolor{gray}{4}&   & \texttt{movw r10, r15} & ;\texttt{HiddenLayerOutput[i]} = \texttt{Accum} \\
\textcolor{gray}{5}&   & \texttt{movw r12, r17} & ;\texttt{HiddenLayerOutput[i]} = \texttt{Accum} \\
\textcolor{gray}{6}&   & \texttt{jmp end} & ;jump after the else statement\\
\textcolor{gray}{7}&  \texttt{else:} & \texttt{clr r10} & ;\texttt{HiddenLayerOutput[i]}$=0$\\
\textcolor{gray}{8}&   & \texttt{clr r11} & ;\texttt{HiddenLayerOutput[i]}$=0$\\
\textcolor{gray}{9}&   & \texttt{clr r12} & ;\texttt{HiddenLayerOutput[i]}$=0$\\
\textcolor{gray}{10}&   & \texttt{clr r13} & ;\texttt{HiddenLayerOutput[i]}$=0$\\
\textcolor{gray}{11}&   \texttt{end:} & ... & ;continue the execution
\end{tabular}
\end{center}
where each float number is stored in $4$ registers.
For example, \texttt{Accum} is stored in registers \texttt{r15,r16,r17,r18} and \texttt{HiddenLayerOutput[i]} is stored in \texttt{r10,r11,r12,r13}.
Line 4,5 executes the equation\\ \texttt{HiddenLayerOutput[i]} = \texttt{Accum}.

The attack was skipping the ``\texttt{jmp end}'' instruction that would normally avoid the part of code setting \texttt{HiddenLayerOutput[i]} to $0$ in case \texttt{Accum} $> 0$.
Therefore, such change in control flow renders the neuron inactive no matter what is the input value.
   
\noindent
\textbf{Sigmoid.} This function was implemented by a following C code:
\begin{center}
   \texttt{HiddenLayerOutput[i] = 1.0/(1.0 + exp(-Accum));} 
\end{center}
After the assembly code inspection, we observed that the successful attack was taking advantage of skipping the negation in the exponent of \texttt{exp()} function, which compiles into one of the two following codes, depending on the compiler version:
\begin{center}
\begin{tabular}{lll}
  A) & \texttt{neg r16} & ;compute negation r16\\
  B) & \texttt{ldi r15, 0x80} & ;load 0x80 into r15 \\
     & \texttt{eor r16, r15} &  ;xor r16 with r15
\end{tabular}
\end{center}
Laser experiments showed that both \texttt{neg} and \texttt{eor} could be skipped, and therefore, significant change to the function output was achieved.

\noindent
\textbf{Hyperbolic tangent.} This function was implemented by a following code in C:
\begin{center}
   \texttt{HiddenLayerOutput[i] = 2.0/(1.0 + exp(-2*Accum)) - 1;} 
\end{center}
Similarly to sigmoid, the experiments showed that the successful attack was exploiting the negation in the exponential function, leading to an impact similar to sigmoid.

\noindent
\textbf{Softmax.} In case of softmax function, we were unable to obtain any successful misclassification.
There were only two different outputs as a result of the fault injection: either there was no output at all, or the output contained invalid values.
This lack of valid output prevented us to do further fault analysis to derive the actual fault model that happened in the device.
Therefore, a thorough analysis of softmax behavior under faults would be an interesting topic for the future work.

\begin{figure}[b]
    \centering
    \setlength{\tabcolsep}{1pt}
    \begin{tabular}{ccc}
    \includegraphics[width=0.31\textwidth]{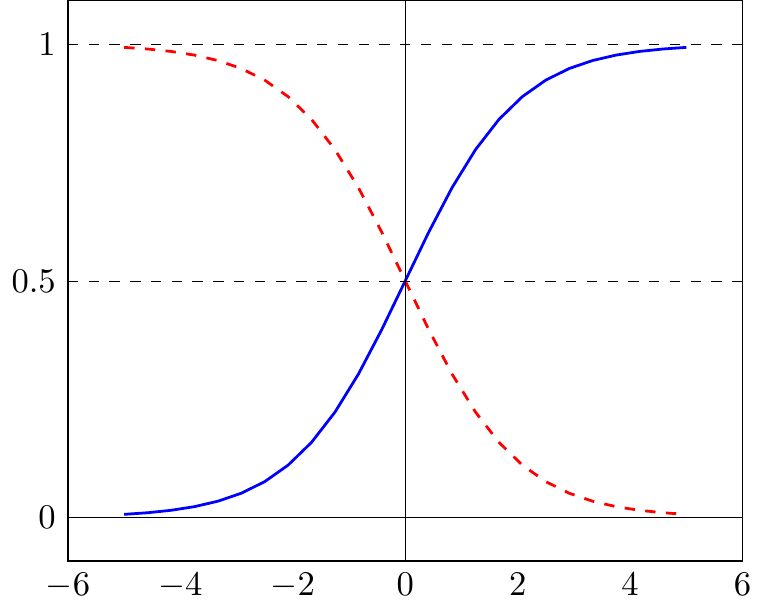}&
    \includegraphics[width=0.31\textwidth]{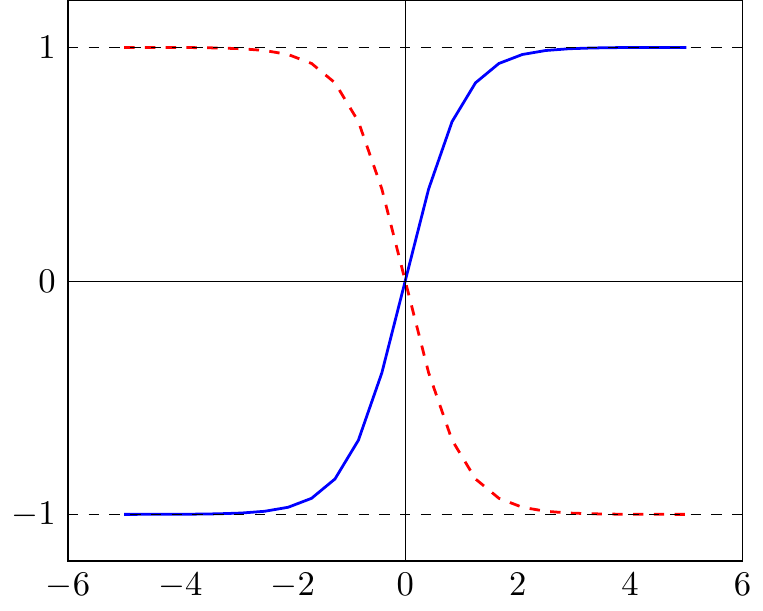}&
    \includegraphics[width=0.31\textwidth]{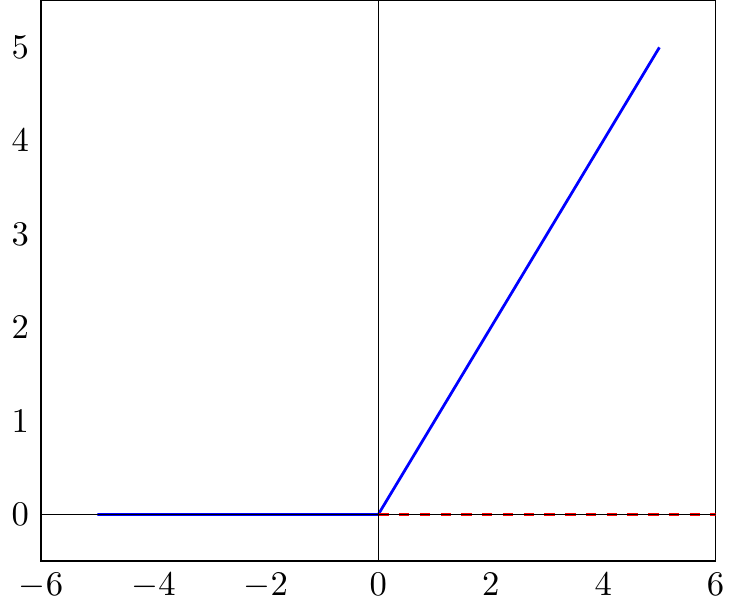}\\
    (a)&(b)&(c)
    \end{tabular}
    \caption{(a) Sigmoid, (b) Hyperbolic tangent, and (c) ReLu functions. Blue lines indicate original function, red lines indicate faulted ones.}
   \label{fig:functions}
\end{figure}

If we let $y$ and $y'$ denote the correct and faulted output of the target activation function, the relation between $y$ and $y'$ is summarized in Table~\ref{tab:attackeffect}. For further illustration, the graph of original and faulted activation functions is depicted in Figure~\ref{fig:functions}.
\begin{table}[tb]
    \centering
    \begin{tabular}{|c|c|}\hline
      Target activation function   & Relation between $y$ and $y'$ \\ \hline\hline
       ReLu  & $y'=0$ \\\hline
       sigmoid & $y'=1-y$ \\ \hline
       tanh & $y'=-y$ \\ \hline
    \end{tabular}
    \caption{Relation between correct output $y$ and faulted output $y'$ when a single fault is injected in target activation function}
    \label{tab:attackeffect}
\end{table}

\section{Application to Deep Neural Network}
The results from previous section aiming at single functions can be directly used to alter the behavior of a neural network.
In this section we extend the attack to a full network, while targeting several function computations at once with a multi-fault injection model.
When it comes to deep neural networks, there are three possible places to introduce a fault:
\begin{itemize}
    \item Input layer -- such fault would be identical to introducing a change at the input data. Therefore, it is of little interest, since it would be normally easier for the attacker to directly alter the input data rather than injecting precise faults with an expensive equipment.
    \item Hidden layer(s) -- since the structure of the hidden layer is normally unknown to the attacker, she cannot easily predict the outcome of the fault injection. However, she can still achieve the missclassification, although not necessarily to the class she decides. Therefore, such attack might be interesting in case the attacker does not care about the outcome class as long as it is different from the correct outcome.
    \item Output layer -- normally, softmax is the function of choice for the output layer. According to our results, introducing a meaningful fault into softmax is harder compared to other functions. However, as we discussed, in case the attacker can alter registers storing the floating point data, she can easily missclasify the outcome to a chosen class, making it a very powerful attack model.
\end{itemize}
One might be wondering how can the attacker develop a strategy if the network is unknown to her.
However, as shown recently~\cite{csi_nn}, it is possible to determine the activation function with a side-channel analysis, i.e. by measuring the power consumption or the electromagnetic emanation from the device during the computation.
Getting the weights and deciding on attacking particular neurons might be trickier, so in this case, a random fault model would be the most reasonable assumption for the attacker.
Deciding on what layer to attack, it makes sense to inject the fault as close to the output layer as possible to make the impact highest.
Therefore, for our case, the attacker randomly injects faults into the 4. hidden layer of the network, targeting multiple activation function computations.
In the following we provide evaluation results for such case.

\noindent
\textbf{Evaluation on a sample DNN.}
To test how our attack can influence a real-world DNN, we trained and evaluated different DNNs with the random fault model described above.
The attack methods considered are described in Section~\ref{sec:attack}.
We have selected a popular MNIST dataset~\cite{lecun1998mnist}.
The training of DNNs was accomplished using Keras (ver.2.1.6)~\cite{chollet2015keras}
and Tensorflow libraries (ver.1.8.0)~\cite{abadi2016tensorflow}.
The structures of the DNNs are detailed in Table~\ref{tab:dnn_structure}.
For each target function (ReLu, sigmoid and tanh), $8$ DNNs with different number of neurons $(n=15,20,30,40,50,60,70,80)$ in hidden layer 4 were evaluated.
We used a partially fixed structure of DNN in order to study the effects of fault attacks on different activation functions.
The prediction accuracy we obtained is summarized in Table~\ref{tab:accuracy}.
The accuracy shows that although the DNNs we choose are relatively simple, their accuracy is comparable with the state of the art.
\begin{table}[tb]
\begin{center}
\begin{tabular}{|c|c|c|}\hline
    Layer & No. of neurons & Activation function \\ \hline\hline
    Input layer & 784 & -\\ \hline
    Hidden layer 1 & 500 & ReLu\\ \hline
    Hidden layer 2 & 500 & ReLu\\ \hline
    Hidden layer 3 & 500 & ReLu\\ \hline
    Hidden layer 4 & n   & target function\\ \hline
    Output layer   & 10  & Softmax \\ \hline
\end{tabular}
\end{center}
\caption{Structure of the DNN used in fault injection simulations.}
\label{tab:dnn_structure}
\end{table}

\begin{table}[h]
\begin{center}
\begin{tabular}{|c|c|c|c|c|c|c|c|c|}\hline
    Target function &
      \multicolumn{8}{c|}{ReLu} \\ \hline
    n & 15 & 20 & 30 & 40 & 50 & 60 & 70 & 80 \\ \hline
    Train. Acc. & 97.8 & 98.9 & 99.5 & 99.1 & 99.2 & 99.4 & 99.2 & 99.1 \\ \hline
    Test.  Acc. & 96.6 & 98.0 & 98.3 & 97.6 & 97.4 & 98.1 & 98.1 & 97.6  \\ \hline\hline
    Target function &
      \multicolumn{8}{c|}{sigmoid} \\ \hline
    n & 15 & 20 & 30 & 40 & 50 & 60 & 70 & 80 \\ \hline
    Train. Acc. & 98.9 & 99.2 & 98.9 & 99.2 & 99.1 & 99.2 & 99.0 & 99.2 \\ \hline
    Test.  Acc. & 97.8 & 97.7 & 97.7 & 98.0 & 98.0 & 98.1 & 97.8 & 98.0  \\ \hline\hline
    Target function &
      \multicolumn{8}{c|}{tanh} \\ \hline
    n & 15 & 20 & 30 & 40 & 50 & 60 & 70 & 80 \\ \hline
    Train. Acc. & 98.6 & 99.2 & 99.1 & 99.0 & 99.0 & 99.1 & 99.2 & 99.0 \\ \hline
    Test.  Acc. & 97.4 & 97.9 & 97.8 & 97.8 & 98.0 & 97.8 & 97.9 & 97.8  \\ \hline
\end{tabular}
\end{center}
\caption{Training/testing accuracy of DNNs used in evaluation.}
\label{tab:accuracy}
\end{table}

\begin{figure}
\centering
\begin{tabular}{cc}
\begin{tikzpicture}[scale = 0.7]
	 \begin{axis}[
	 		xlabel= \textbf{No. of neurons in the layer},
	 		ylabel= \textbf{Success rate of missclassification},
	 		legend pos={north west},
	 		ymin=0,ymax=100,
	 		xtick={15, 20, 30, 40, 50, 60, 70, 80, 90},
	 		]
	 \addplot[red,thick] plot coordinates{
	 (15, 20.38)
	 (20, 9.63)
	 (30, 2.25)
	 (40, 0.88)
	 (50, 0.75)
	 (60, 0.00)
	 (70, 0.00)
	 (80, 0.13)
	 };\addlegendentry{ReLu}
     
     \addplot[green,dashdotted,thick] plot coordinates{
     (15, 64.00)
     (20, 13.38)
     (30, 1.25)
     (40, 1.00)
     (50, 0.38)
     (60, 0.50)
     (70, 0.25)
     (80, 0.38)
	 };\addlegendentry{Sigmoid}
	 
	\addplot[blue,dashed,thick] plot coordinates{
     (15, 61.25)
     (20, 14.12)
     (30, 1.13)
     (40, 0.75)
     (50, 0.50)
     (60, 0.75)
     (70, 0.50)
     (80, 0.13)
	 };\addlegendentry{tanh}
\end{axis}
\end{tikzpicture}
&
\begin{tikzpicture}[scale = 0.7]
	 \begin{axis}[
	 		xlabel= \textbf{No. of neurons in the layer},
	 		ylabel= \textbf{Success rate of missclassification},
	 		legend pos={north east},
	 		ymin=0,ymax=100,
	 		xtick={20, 30, 40, 50, 60, 70, 80, 90},
	 		]
	 \addplot[red,thick] plot coordinates{
	 (20, 57.50)
	 (30, 19.25)
	 (40, 3.00)
	 (50, 1.00)
	 (60, 1.13)
	 (70, 0.13)
	 (80, 0.38)
	 };\addlegendentry{ReLu}
     
     \addplot[green,dashdotted,thick] plot coordinates{
     (20, 100.00)
     (30, 98.88)
     (40, 52.00)
     (50, 11.25)
     (60, 1.75)
     (70, 0.88)
     (80, 0.38)
	 };\addlegendentry{Sigmoid}
     
     \addplot[blue,dashed,thick] plot coordinates{
     (20, 100.00)
     (30, 98.13)
     (40, 48.38)
     (50, 9.50)
     (60, 1.75)
     (70, 0.88)
     (80, 1.75)
	 };\addlegendentry{tanh}
\end{axis}
\end{tikzpicture}\\
(a) & (b)
\end{tabular}
\caption{Injecting (a) 5 and (b) 15 random faults in hidden layer 4.}
\label{fig:random}
\end{figure}
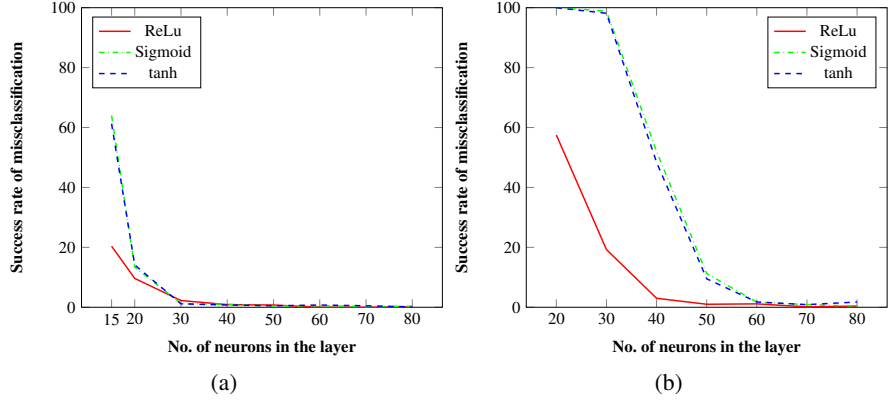

Figures~\ref{fig:random} (a) and~\ref{fig:random} (b) show a random fault model when attacking the last hidden layer of the network with 5 and 15 faults, respectively.
Success rates are calculated for $800$ random inputs.
Naturally, with increasing number of neurons in the layer, the success rate drops.
The figures also show that sigmoid and tanh functions follow the same trend, which is caused by the same type of fault as explained in the previous section -- skipping the negation in the exponentiation function.

Overall, it can be concluded that if the attacker wants to have a reasonable success rate (>50\%), she should inject faults in at least half of the neurons in the chosen layer, in case of sigmoid and tanh.
For ReLu, she should fault at least 3/4 of the neurons.

\section{Conclusion and Future Work}
In this paper, we have proposed the first physical fault injection attack technique on the major activation functions of deep neural networks.
We stated implications how such attack can alter the behavior of targeted network, together with simulations.
Our results demostrate practicality of the attack on ReLu, sigmoid, and tanh.

For softmax, assuming that we could use single bit flip on each register during the execution~\cite{agoyan2010flip}, we could target a specific misclassification of the label. 
Assuming an 8-bit microcontroller, the floating point data is represented using IEEE 754 standard.
Since the target device is an 8-bit microcontroller,  the representation follows 32-bit pattern $(b_{31}...b_{0})$, which is stored in 4 registers.
The 32-bits consist of: 1 sign bit $(b_{31})$, 8 biased exponent bits $(b_{30}...b_{23})$ and 23 mantissa (fractional) bits $(b_{22}...b_{0})$. It can be formulated as: $(-1)^{b_{31}}\times 2^{(b_{30}...b_{23})_2 - 127}\times (1.b_{22}...b_{0})_2$.
For example, the value $2.43$ can be expressed as $(-1)^{0}\times 2^{(1000000)_2 - 127}$ $\times (1.00110111000010100011111)_2$.
Since 32-bit $m$ is split into 4 bytes, each byte of $m$ is targeted individually.
Hence, one can target the exponent byte on the output neuron for specific label. For example, from the example before, the first byte will contain $01000000$.
If bit flip is performed on sixth bit, the resulting byte will be $01000100$ and the corresponding floating point value will be $622.08$, and hence, during softmax computation, the probability for the targeted class will increase (assuming other values are kept the same) and it will force the algorithm to classify the data to this class.

This area is still in the beginning phase of research and therefore, it is unclear what practical consequences can happen when such attack is aimed for example at autonomous vehicle system or some other critical expert system.
While the laser fault injection, used in our experiments, might not be possible in many scenarios, there are much simpler ways to disturb the circuits, such as voltage/clock glitching or electromagnetic fault injection~\cite{bar2006sorcerer}.
In the future, we will also explore such applications.

It will also be interesting to look at possible countermeasures.
While there are already techniques available that correct non-malicious alterations of the processed values in DNN (due to environmental conditions)~\cite{lee2014fault}, the fault tolerance techniques against malicious entities have to be developed in the same way as in the area of applied cryptography~\cite{breier2017feeding,servant2014study,ciet2005practical}.


\end{document}